\documentclass[pra,aps,twocolumn,showpacs,amsmath,amssymb]{revtex4}
\newcommand{\eins}{I}
\newcommand{\ket}[1]{ |#1  \rangle}
\newcommand{\bra}[1]{ \langle #1|}
\newcommand{\proj}[1]{\ket{#1}\bra{#1}}
\newcommand{\dket}[1]{ | #1  \rangle\!\rangle}
\newcommand{\dbra}[1]{ \langle\!\langle #1|}
\newcommand{\dproj}[1]{\dket{#1}\dbra{#1}}
\def\){\rangle\!\rangle}\def\({\langle\!\langle}
\def\set#1{{\cal #1}}\def\sH{\set{H}}
\def\transp#1{{#1}^T}\def\partransp#1{{#1}^\theta}
\def\d{\operatorname{d}}\def\<{\langle}\def\>{\rangle}
\def\Tr{\operatorname{Tr}}
\begin{document}
\title{Local observables for entanglement witnesses}
\author{G. Mauro D'Ariano}
\email{dariano@unipv.it}
\author{Chiara~Macchiavello}
\email{chiara@unipv.it}
\author{Matteo G.A. Paris}
\email{paris@unipv.it}
\affiliation{{\em Quantum Optics and Information Group}, Istituto
Nazionale di Fisica della Materia, Unit\`a di Pavia}
\homepage{http://www.qubit.it}
\affiliation{Dipartimento di Fisica ``A. Volta", Universit\`a di Pavia, Italy}
\date{\today}
\begin{abstract}
We present an explicit construction of entanglement witnesses for depolarized
states in arbitrary finite dimension. For infinite dimension we generalize the
construction to twin-beams perturbed by Gaussian noises in the phase and in
the amplitude of the field. We show that entanglement detection for all these
families of states requires only three local measurements.  The explicit form
of the corresponding set of local observables (quorom) needed for entanglement
witness is derived.
\end{abstract}
\maketitle
\section{Introduction}
Entanglement plays an essential role in almost all aspects of quantum
information theory \cite{nielsen}. Entangled states are the key ingredients of
many quantum protocols such as quantum teleportation, quantum dense coding,
and entanglement-based quantum cryptography. However, entanglement can be in
general corrupted by the interaction with the environment. Therefore,
entangled states that are available for experiments are usually mixed states,
and it becomes crucial to establish whether or not entanglement has survived
the environmental noise. 
\par
The issue of experimental entanglement detection was first addressed for pure 
states in Ref. \cite{susana}. More recently, 
in Ref. \cite{eh} procedures based on the use of collective measurements were
proposed. Later, in Ref. \cite{exdet} a general method to detect entanglement 
with few local measurements was presented and optimal schemes were designed for
two-dimensional systems, bound entangled states and entangled states
of three qubits.
In Ref. \cite{rubin} a method for local detection of nonseparable states has
been derived for bipartite states in dimension $d$ and to some families of 
states of $n$ qubits; it was shown in particular that in the bipartite case
and for $d$ a prime number the method achieves the lower bound of $d+1$ 
measurements derived in Ref. \cite{exdet}.
In this paper we extend the approach of \cite{exdet} to depolarised bipartite
states in arbitrary dimension, and show how entanglement can be 
efficiently detected by identifying the minimal needed set of local
observables, the so-called {\em quorum} of observables.  
Moreover, we address the problem of entanglement detection
for continuous variables (CV) and find entanglement witnesses (EW) for 
a twin-beam state (TWB) corrupted by Gaussian noises, both in the
phase and in the amplitude of the field. In this case efficient
homodyne-tomographic procedures are analyzed suited to
local detection of entanglement. 
We found that for all the families of states that we have considered a 
rank-four witness operator is sufficient to detect entanglement.
Notice that this result is not in contradiction with the ones derived in 
Ref. \cite{rubin} because we assume to have more knowledge about the
family of states.
\par
The paper is organized as follows. In Sect. \ref{s:d} we construct the
EW for bipartite depolarized entangled states 
in arbitrary finite dimension, and give the explicit
form of the corresponding local quorum. 
In Sect. \ref{s:inf} we analyze the case of bipartite CV systems. 
In particular we study the family of twin beam states corrupted
by Gaussian noise, both in the phase and in the amplitude of the
field, and show how to detect entanglement by employing homodyne 
tomographic techniques. In Sect. \ref{conc} we close the paper with 
a summary of the results and final comments.
\section{Depolarized states in arbitrary dimension}
\label{s:d}
In this section we will show how to detect entanglement locally for
depolarized states in arbitrary finite dimension $d$, namely for the family
of states
\begin{equation}
\rho=p\proj{\psi}+\frac{1-p}{d^2}\eins\otimes \eins\;,
\label{defW}
\end{equation}
where $\ket{\psi}$ is any bipartite entangled normalized 
pure state of systems with dimension $d$, 
$\eins$ is the $d\times d$ identity operator and $0\leq p\leq 1$. 
If $\ket{\psi}$ is a maximally entangled state, the states in Eq.
(\ref{defW}) coincides with the family of the socalled isotropic states. 
\par
We will now introduce a more convenient notation. 
Given a bases
$\{\ket{i}\otimes\ket{j}\}$ for the Hilbert space 
$\sH_1\otimes\sH_2$ (with $\sH_1$ and $\sH_2$ generally not isomorphic), we can write any vector $\dket{\Psi}\in \sH_1
\otimes \sH_2$ as
\begin{equation}
\dket{\Psi}=\sum_{ij} \Psi_{ij}\ket{i}_1\otimes\ket{j}_2\;.
\label{defC}
\end{equation}
The above notation \cite{bellobs} exploits the correspondence between states $\dket{\Psi}$ 
in $\sH_1\otimes \sH_2$ and
Hilbert-Schmidt operators $\Psi=\sum_{ij}\Psi_{ij}\ket{i}\bra{j}$from
$\sH_1$ to $\sH_2$.
The following relations are an immediate consequence of the definition
(\ref{defC}):
\begin{eqnarray}
&& A\otimes B\dket{\Psi}=\dket{A\Psi\transp{B}}\;,\\
&&\(A\dket{B}=\Tr[A^\dagger B]\;,
\label{defop}
\end{eqnarray}
where $\transp{B}$ denotes the transposition of the operator $B$
with respect to the chosen basis $\{|i\>\}$.
As mentioned above, in the following we will consider only bipartite states on 
$\sH\otimes \sH$, where $\sH$ has dimension $d$.
\par
In this notation the depolarized  state (\ref{defW}) takes the form
\begin{equation}
R=p\dproj{\Psi}+\frac{1-p}{d^2}\eins\otimes\eins\;.
\label{defW2}
\end{equation}
Let us briefly recall the definition of
EW \cite{terwit,opti}. A state $\rho$ is entangled iff 
there exists an Hermitian operator $W$ such that $\Tr[W\rho]<0$, 
while $\Tr [W\rho_{sep}]\geq 0$ for all separable states $\rho_{sep}$.
The operator $W$ is called {\em entanglement witness} (EW).
For entangled states with non positive partial transpose (NPT) $W$ can be 
explicitly constructed as $W=\partransp{(\proj{\epsilon})}$, where
$\partransp{O}$ denotes the partial transposed of $O$ on the second
Hilbert space, and $\ket{\epsilon}$
is the eigenvector of $\partransp{\rho}$ that corresponds to the minimum 
eigenvalue \cite{opti}.
Notice that this is not the only method to construct entanglement witnesses.
Other techniques, working for both NPT and PPT entangled states, have been
suggested, as for example, in Refs. \cite{doherty,hradil}.
\par
The entangled states of the form (\ref{defW}) 
have non positive partial transpose \cite{reduction,vidal}.
Following the approach of \cite{exdet}, we will show how to detect
entangled states within the family (\ref{defW})
by explicitly deriving EW according to the above
construction.
\par
The partial transpose of the state $R$ can be written as 
\begin{equation}
\partransp{R}=p(\Psi\otimes \eins) E (\Psi^\dagger\otimes\eins)
+\frac{1-p}{d^2}\eins\otimes\eins\;,
\label{ptR}
\end{equation}
where $E$ is the swap operator, i.e. $E=\sum_{ij}\ket{i}\bra{j}\otimes 
\ket{j}\bra{i}$. 
\par
As mentioned above, in order to construct a  witness operator for the family 
of states (\ref{defW}), we look for the eigenvector of $\partransp{R}$ corresponding
to the minimum eigenvalue. Therefore, we can start by writing explicitly the
eigenvalue equation
\begin{equation}
\partransp{R}\dket{A}=\lambda\dket{A}\;,
\label{eigen}
\end{equation}
where $\dket{A}$ is the eigenvector for the eigenvalue $\lambda$.
By using the properties (\ref{defop}) and Eq. (\ref{ptR}), we can also write
\begin{equation}
\partransp{R}\dket{A}=p\dket{\Psi\transp{A}\Psi^*}+c\dket{A}\;,
\label{ptR2}
\end{equation}
where $c=(1-p)/d^2$, and $O^*$ denotes complex conjunction of the
operator $O$ 
with respect to the chosen basis $\{|i\>\}$.
Therefore, the eigenvalue equation in operatorial terms takes the form
\begin{equation}
\lambda A=p\Psi\transp{A}\Psi^*+cA\;,
\label{eqav}
\end{equation}
and can be more conveniently written as
\begin{equation}
\Psi\transp{A}\Psi^*=\mu A,\qquad\mu=(\lambda -c)/p\;. 
\label{eqav2}
\end{equation}
\par
We now use the singular value decomposition of the matrix
$\Psi$, namely $\Psi=X\Sigma Y^\dagger$, where $X$ and $Y$ are 
unitary operators, while 
$\Sigma$ is the diagonal operator containing the eigenvalues 
$\{\sigma_j\}$ of
$\sqrt{\Psi\Psi^\dagger}$---the so-called {\em singular values}
of $\Psi$---which are conventionally ordered decreasingly. 
The above equation then takes the form
\begin{equation}
X\Sigma Y^\dagger\transp{A} X^* \Sigma \transp{Y}=\mu A\;.
\label{eqav4}
\end{equation}
By multiplying Eq. (\ref{eqav4}) by $X^\dagger$ on the left and by $Y^*$ 
on the right, and upon
defining 
\begin{equation}
B=Y^\dagger\transp{A} X^*,\label{defB}
\end{equation}
Eq. (\ref{eqav4}) can be written 
in the compact form
\begin{equation}
\transp{B}=\mu^{-1} \Sigma B\Sigma\;.
\label{B}
\end{equation}
The last equation can be 
conveniently expressed by explicitly writing its matrix elements as follows
\begin{equation}
b_{ij}=\mu^{-1}b_{ji}\sigma_i\sigma_j\:.
\label{bij}
\end{equation}
By reiterating the above equation one obtains
\begin{equation}
b_{ij}=\mu^{-2}\sigma^2_i\sigma^2_j b_{ij},
\label{bij2}
\end{equation}
which is fulfilled for 
\begin{equation}
\mu^2=\sigma^2_i\sigma^2_j.\label{fulfill}
\end{equation}
For values of $i$ and $j$ that cannot satisfy Eq. (\ref{fulfill}) we necessarily have $b_{ij}=0$. 
We now want to specify the 
form of the operator $B$ corresponding to the minimum eigenvalue $\lambda$.
Notice first that for eigenvalues $\lambda<c$ the parameter $\mu$ is negative, 
and therefore, according to Eq. (\ref{bij}), all diagonal elements of $B$
vanish. This is the case in particular when the minimum eigenvalue $\lambda_m$
is negative.
We will now explicitly derive the form of $B$ corresponding to the minimum
eigenvalue $\lambda_m$. Suppose that $\sigma_1$ and $\sigma_2$ are the two
largest elements of $\Sigma$  and $\sigma_1\geq\sigma_2$. Then, from
Eq. (\ref{eqav2}) the minimum 
eigenvalue $\lambda_m$ takes the form
$\lambda_m=-p\sigma_1\sigma_2+c$, and according to Eq. 
(\ref{bij2}) the matrix elements of the operator $B$ corresponding to
$\lambda_m$ (which we will denote by $\bar B$) are
\begin{equation}
\bar b_{12}=-\bar b_{21}=1\;,
\label{barb}
\end{equation}
while all the other elements vanish. Therefore, the operator $\bar B$ has 
rank two and takes the explicit form
\begin{equation}
\bar B= \left( \begin{array}{c|cc} 
\begin{array}{cc} 0 & 1 \\ -1 & 0 \end{array} & &\hbox{\Large 0} \\
\hline & \\ 
\hbox{\Large 0} & &\hbox{\Large 0} \end{array}\right)
\label{Bmat}
\end{equation}
The expression for the operator $A$ corresponding to the minimum 
eigenvalue $\lambda_m$, which we will call $\bar A$, follows from the 
definition of $B$ in Eq. (\ref{defB}) and is given by
\begin{equation}
\bar A=X\bar B\transp{Y}\;.
\label{barA}
\end{equation}
The EW for the family of states (\ref{defW}) can then be 
derived as  
\begin{equation}
\bar W=\partransp{(\dproj{\bar A})}.\label{WA}
\end{equation}
Notice that the same form of $\bar B$ is valid also for
degenerate maximum singular value $\sigma_1$, although in this case
the solution  is not unique. 
Moreover, an interesting feature of the resulting witness operator is that 
its rank is four, independently of the dimension $d$ of the
subsystems. We also want to point out that the EW for
the states (\ref{defW}) does not depend on the value of $p$, but only on some
a priori knowledge about the state $\dket{\Psi}$, namely on the singular
values of $\Psi$ and on the form of the operators $X$ and $Y$.
\par
As an illustration we will consider two explicit examples. When $\dket{\Psi}$
is a maximally entangled state in dimension $d$ of the form
$\dket{\Psi}=\tfrac{1}{\sqrt d}\sum_j\ket{jj}$, i.e. the operator $\Sigma$ is 
proportional to the identity, with $\sigma_i=1/\sqrt{d}$, 
then the operator $\bar A$ 
corresponding to a state $\dket{\bar A}=(\ket{ij}-\ket{ji})/\sqrt{2}$ can be
used to construct a witness operator. In this case 
the state is separable iff $p>1/(d+1)$. \par
As a second example let us consider an initial state with Schmidt number two,
i.e. $\sigma_1=\sigma_2=1/\sqrt{2}$ and $\sigma_i=0$ for $i>2$.
In this case the corresponding EW is constructed from
$\dket{\bar A}=(\ket{01}-\ket{10})/\sqrt{2}$, where $\ket{01}$ and  
$\ket{10}$ are the basis states related to $\sigma_1$ and $\sigma_2$.
The state is entangled when $p\geq 2/(d^2+2)$.
\par
We will now show how to detect entanglement for the family of states 
(\ref{defW}) by measuring only three local observables. The matrix $\bar A$ 
in (\ref{barA}) can be written as 
\begin{equation}
\bar A = i X \:(\sigma_y \oplus {\bf 0})\: \transp{Y}\;,
\end{equation}
where $\sigma_y$ is a Pauli matrix (acting between the two levels of the 
two-dimensional subspace spanned by $\bar A$), $\oplus$ denotes the direct sum, 
and ${\bf 0}$ is the null matrix. If $P$ is the projection operator over the 
subspace where $\bar A$ is not null, the above expression can be rewritten as 
\begin{equation}
\bar A = i X^\prime P^\prime \Sigma_y P^\prime Y^*\:, 
\end{equation}
where $X^\prime=X\transp{Y}$, $P^\prime=Y^* P\transp{Y}$, and 
$\Sigma_y=Y^* \sigma_y \oplus {\bf 0}\transp{Y}$.
Inserting the above expression in the definition (\ref{WA}) 
of $\bar W=(\bar A\otimes \eins) 
E (\bar A^\dagger \otimes \eins)$ we have
\begin{equation}
W= (X^\prime \Sigma_y \otimes \eins) (E_2 \oplus {\bf 0})
(\Sigma_y X^{\prime\dag}\otimes \eins)\:,
\end{equation}
where $E_2$ is the swap operator for the two-dimensional subspace
spanned by the support of 
$\bar A$. Since one has 
\begin{equation}
E_2=\tfrac{1}{2}\sum_{\alpha=t,x,y,z}\sigma_\alpha\otimes \sigma_\alpha,
\end{equation}
where $\sigma_t\equiv I$, the EW can be finally
written as 
\begin{equation}
\bar W =\tfrac{1}{2}  I\otimes I+
\sum_{\alpha=x,y,z}\tfrac{1}{2}
\tilde\sigma_\alpha\otimes\sigma_\alpha
\label{EWf}\;,
\end{equation}
with
\begin{equation}
\tilde\sigma_\alpha=
X^\prime \Sigma_y \sigma_\alpha\Sigma_y X^{\prime\dag}. 
\end{equation}
As we can see from Eq.(\ref{EWf}), the witness operator $\bar W$ can
be measured by performing the measurements of only three local 
observables $\tilde\sigma_\alpha\otimes\sigma_\alpha$, $\alpha=x,y,z$.
This result generalizes that of Ref. \cite{exdet} to arbitrary
dimension for states of the form (\ref{defW}): in all cases only three
local observables are sufficient.
\par
As mentioned in the introduction, in Ref. \cite{rubin} a different method to detect 
entanglement of $d$ dimensional states has been proposed.
This method is valid for states of the form $\ket{\psi}=\sum_{k=0}^{d-1}
a_k\ket{kk}$ with $a_k\geq 0$ and requires the measurement of $d+1$ 
observables.
Compared to our method, it needs the measurements of a larger number of
observables, but, on the other hand, it does not require the knowledge of the 
values of the coefficients $a_k$ in the density matrix. 
\section{Perturbed twin-beam in continuous variables}
\label{s:inf}
In this section we address the construction and the measurement
of EW for CV. At
first we have to define the families of states we are going to
consider. These cannot be a trivial generalization of the isotropic
states, since both maximally entangled states and the identity
are unphysical states in an infinite dimensional Hilbert space.
We start from the ``maximally'' entangled state of two CV systems at
finite energy, which is given by
\begin{equation}
|\Psi\)=\Psi\otimes I |I\),\quad\Psi=\sqrt{1-|x|^2}e^{-xa^\dag a},\; |x|<1\label{twb},
\end{equation}
where without loss of  generality we will consider $x$ as real. Here
and in the following,  with $a^\dag$, $b^\dag$, and
$a$, $b$ we will denote the creation and annihilation
operators of two independent harmonic oscillators, respectively, with
commutations $[a,a^\dag]=[b,b^\dag]=1$. For the e. m. radiation the
harmonic oscillators describes two field modes, and
 Eq. (\ref{twb}) describes the so-called twin-beam state 
(TWB) obtained by parametric downconversion of the vacuum in a 
nondegenerate optical parametric amplifier. In this case $\bar{n}=2x^2/(1-x^2)$
represents the average number of photons of the TWB.
In practice, TWB are the most reliable source of CV entanglement: indeed, 
experimental implementation of quantum information protocols such as teleportation,
have been obtained using TWB of radiation. 
\par
Let us now analyze the family of states that are obtained by
perturbing a TWB with a noisy environment. We will consider
Gaussian noises both in the phase and in the amplitude of the field
modes. Thermal noise is a special case of the 
present Gaussian displacement noise, whereas the noise coming from 
the addition of a thermal state has been considered in \cite{fiura}.
In this case our results coincide with the ones given there.
\par
The action of a phase-destroying environment on the TWB 
is described by the  Master equation 
\begin{eqnarray}
\dot{R}&=&\tfrac{\gamma}{2}\left[
2a^\dag aR a^\dag a 
-(a^\dag a)^2R -R (a^\dag a)^2 
\right.\\ &+& \left. 2b^\dag b
R b^\dag b-(b^\dag b)^2R-R
(b^\dag b)^2\right] 
\label{phaseD}\;,
\end{eqnarray}
where $\dot{R}$ denotes the time derivative of the state $R$.
The solution of Eq. (\ref{phaseD}) for initial condition $R_0 =
\dproj{\Psi}$, can be expressed as
\begin{equation}
R(t) = (1-x^2) \sum_{p,q} \: x^{p+q} \: e^{-\gamma t
\left|p-q\right|^2}\: |pp\rangle\langle qq|
\label{phase-family}\;,
\end{equation}
where we used the abbreviate notation $\ket{ij}$ for $\ket{i}\otimes\ket{j}$. The correlations between the modes are reduced in the mixture
(\ref{phase-family}) compared to the initial TWB state. However, 
as we will see by explicitly constructing an EW, phase-noise never leads
to a separable state, {\em i.e.} the entanglement is not destroyed for any
value of $\gamma t$.
\par
In order to obtain an EW for the family $R(t)$ we 
construct and diagonalize the partial transpose $\partransp{R}(t)$
\begin{equation}
\partransp{R}(t) = (1-x^2) \sum_{pq} \: x^{p+q} \: e^{-\gamma t
\left|p-q\right|^2}\: |pq\rangle\langle qp|.
\label{twbgmtheta}
\end{equation}
The eigenvalues equation $\partransp{R_\gamma} \: |\psi\)=\lambda
|\psi\)$ is solved by
\begin{eqnarray}
\lambda_n &=& (1-x^2) x^{2n},\qquad |\psi_n\rangle=|nn\rangle\;,\nonumber \\
 \lambda_{nm}^{\pm} &=& \pm (1-x^2) x^{n+m} e^{-\gamma t(n-m)^2},\label{eigentwbgm}\\
|\psi_{nm}^{\pm}\)&=&
\tfrac{1}{\sqrt{2}} ( |nm\rangle
\pm |mn\rangle)\nonumber
\end{eqnarray}
The minimum eigenvalue is given by $\lambda_{01}^-=-(1-x^2) x e^{-\gamma}$
corresponding to the eigenvector 
\begin{equation}
|\psi_{01}^-\)= \tfrac{1}{\sqrt{2}}(
|01\rangle - |10\rangle)\;.
\end{equation}
The eigenvector $|\psi_{01}^-\)$ does not depend on
$\gamma t$, and thus is suitable to build a proper EW for this family 
of states. We have
\begin{eqnarray}
W&=&\tfrac12 \partransp{(|\psi_{01}^- \rangle\rangle
\langle\langle\psi_{01}^- |)}\label{EWgamma}
\\ &=&\tfrac12 \left(
|01\rangle\langle 01| +|10\rangle\langle 10| -
|00\rangle\langle 11| -|11\rangle\langle 00|
\right).\nonumber
\end{eqnarray}
The expectation value
\begin{equation}
\Tr\left[R(t)\: W\right]= \lambda_{01}^- < 0 \quad \forall
\,t, x 
\label{expectW}\;
\end{equation}
is always negative and thus the state $R(t)$ is never separable, 
for any value of $t$, and for any value of the initial TWB 
parameter $x$. In other words, although decreased the entanglement is never 
destroyed by phase-noise. It can also be proved \cite{bin} that 
$R(t)$ can be distilled. The result in Eq. (\ref{expectW}) proves the 
conjecture suggested in \cite{hiroshima}, where the entanglement analysis 
of a phase-perturbed TWB was performed by numerical evaluation of 
the relative entropy of entanglement.
\par Let us now consider the family of states obtained by perturbing a
TWB state by Gaussian amplitude noise, namely
\begin{equation}
R_\kappa ={\cal{G}}_\kappa\otimes{\cal{G}}_\kappa(
|\Psi\)\(\Psi|)\label{GaussianNoise},
\end{equation}
where for a single mode state $\rho$ one the map of the Gaussian noise
is given by 
\begin{equation}
{\cal{G}}_\kappa(\rho)\doteq\int\frac{\d^2\alpha}{\pi\kappa}
e^{-\tfrac{|\alpha|^2}{\kappa}}
D(\alpha)\rho
D^\dag (\alpha),
\end{equation}
$D(\alpha)=\exp\{\alpha a^\dag - \bar\alpha a\}$
denoting the displacement operator. We notice that the operator 
(\ref{EWgamma}) obtained for phase-perturbation is an EW also for Gaussian amplitude
noise. Omitting positive factors, we have 
\begin{equation}
\Tr\left[R_\kappa \: W\right] \propto 
\kappa -1 + \frac12 \frac{1-x}{1+x} \stackrel{\bar{n}\gg 1}{\simeq} 
\kappa -1 + \frac1{4\bar{n}}
\label{expectWGaussian}\;.
\end{equation}
Eq. (\ref{expectWGaussian}) says that $R_\kappa$ becomes
separable if $\kappa\gtrsim 1 - \frac14  \bar{n}^{-1} $, a result that
can be also obtained by direct check of the positivity of the
partial transpose (PPT condition) \cite{ent_meas}. The family
$R_\kappa$, in fact, is composed of Gaussian states, for
which the PPT condition is necessary and sufficient for
separability \cite{simon}. It should be mentioned that the
constructive procedure suggested in Ref. \cite{opti} fails to
provide an EW for the the family $R_\kappa$, in particular
it does not lead to a state-independent witness.  
\par
In principle, the EW (\ref{EWgamma}) can be measured by using only three
observables, as in the finite dimensional case. However, there is no
feasible implementation of the measuring apparatus corresponding to
the quorum in the present CV case. Since we are interested only in the
expectation value of $W$, we could use quantum tomography
(for a recent tutorial review on quantum tomography see
Ref. \cite{tomo_lecture}). However, a tomographic determination of 
$W$ is useful only if requires a smaller number of observables than 
those needed for reconstructing the full state. Indeed, this is the 
case for the EW in Eq. (\ref{EWgamma}). In fact, for two modes of
radiation $a_1$ and $a_2$, the expectation value
$\<O\>\doteq\Tr\left[RO\right]$ of a generic
operator $O$ can be obtained by {\em local} repeated measurements
of the quadratures $X_{1\phi_1}=\frac{1}{2}(a_1^\dag e^{i\phi_1}+a_1
e^{-i\phi_1})$ and $X_{2\phi_1}=\frac{1}{2}(a_2^\dag e^{i\phi_2}+a_2
e^{-i\phi_2})$ as follows
\begin{equation}
\< O\>=\iint
\frac{d\phi_1}{\pi} \frac{d\phi_2}{\pi}\<
R[O](X_{1\phi_1},\phi_1;X_{2\phi_2}\phi_2)\>
\label{twomodeQT}\;,
\end{equation}
namely by averaging the over the phases $\phi_{1,2}$
and over an ensemble of repeated measurements
the function of the two quadratures 
$R[O](x_1,\phi_1;x_2,\phi_2)$---so-called {\em estimator}
or {\em kernel function}---depending on the operator $O$. 
The kernel function for Hilbert-Schmidt operators can be obtained 
directly by means of the trace \cite{tomo_lecture} 
$R[O](x_1,\phi_1;x_2,\phi_2)=\Tr
\left[R(X_{1\phi_1}-x_1)R(X_{2\phi_2}-x_2)O\right]$ with
$R(x)=-\lim_{\varepsilon \rightarrow 0^+} 
\frac{1}{2} \hbox{Re} (x+i\varepsilon)^{-2}$.
For the operator $W$ in Eq. (\ref{EWgamma}) we have
\begin{eqnarray}
&&R[W](x_1,\phi_1;x_2,\phi_2) = 
f_{00}(x_1)f_{11}(x_2)\label{kernelW}\\ && + f_{11}(x_1)f_{00}(x_2)- 2 \cos(\phi_1+\phi_2) 
f_{01}(x_1)f_{01}(x_2)\;,\nonumber
\end{eqnarray}
where 
\begin{eqnarray}
f_{00}(x)&=&2\: \Phi(1,\tfrac12;-2x^2) \nonumber\\ f_{01}(x)&=& 4\sqrt{\pi}\: x\: 
\Phi(2,\tfrac32;-2x^2)\label{spefun} \\ 
f_{11}(x)&=&2 \left[ \Phi(1,\tfrac12;-2x^2) -2 \Phi(2,\tfrac12;-2x^2)\right]\;,  \nonumber
\end{eqnarray}
and $\Phi(a,b;z)$ denotes the confluent hypergeometric function.
Remarkably, $R[W]$ depends only on the sum of the two phases
$\phi_{1,2}$, and shows only a couple of oscillations. Therefore,
the number of measurements to detect the entanglement witness is much
smaller than that needed to reconstruct just the first few matrix
elements of the state, say, in the photon number representation,
since the number of oscillations of the estimators for such matrix
elements increases linearly with their photon-number index.
The precision of the tomographic estimation can be further improved by
adaptive techniques \cite{adapt}. 
\par If we are allowed to mix the two modes after the
perturbation, the characterization of entanglement for the family
$R_\kappa$ can be obtained by measuring a single quadrature.
In fact, for Gaussian states, a necessary and sufficient condition 
to have entanglement after a beam splitter is that the two inputs show 
squeezing (in mutually orthogonal directions) \cite{visent,wang}. 
Therefore, if we impinge the two modes of the perturbed TWB in a 
beam splitter, and then measure the quadrature $X=
\frac{1}{2}(a^\dag +a)$ on 
the sum mode, we have squeezing if and only if the input state
is entangled. Therefore, the fluctuation operator $W=\Delta
X^2 -1/4=X^2 - \<X\>^2 - 1/4$ is an EW, and its expectation value is of
course obtained by measuring the quadrature $X$. The analysis is valid 
also when the TWB initial parameter $x$ is complex, in which case 
the phase of the quadrature to be measured coincides 
with the phase of $x$. 
Obviously, if the mixing of the two modes is not possible, one can
always reconstruct the above quadrature locally
by quantum tomography.
\section{Conclusions}\label{conc}
In this paper we have given an explicit construction of EW for depolarised 
states in arbitrary finite dimension. For infinite dimensions,
i. e. for CV, we have
introduced isotropic states as twin-beams perturbed by Gaussian noises
in the phase or in the amplitude of the field, and we have constructed
their respective EW as well. We have shown that in all cases
entanglement detection needs only a quorum of three local observables,
whose explicit form have been derived. For CV it is possible to
use also homodyne tomography efficiently to detect entanglement, 
without determining the matrix elements of the state. 
\bigskip\acknowledgements
We wish to thank Dagmar Bru\ss $\;\;$ for useful comments, and Oliver Rudolph
for pointing out Ref. \cite{vidal}.   This work has been supported by the INFM
project PRA-2002-CLON, by the MIUR project {\em Entanglement assisted high
precision measurements}, and by the EC project EQUIP (IST-1999-11053).  MGAP
is research fellow at {\em Collegio Alessandro Volta}.

\end{document}